\documentstyle[12pt]{article}
\newcommand{\nc}{\newcommand}
\nc{\Ec}{{\varepsilon}} \nc{\Lc}{{\cal L}}  \nc{\munu}{{\mu\nu}}
\nc{\pp}{/\!\!\!p} \nc{\pipi}{/\!\!\!\pi}
\nc{\bb}{\begin{equation}} \nc{\ee}{\end{equation}}
\nc{\bega}{\begin{eqnarray}} \nc{\ega}{\end{eqnarray}}
\nc{\begae}{\begin{eqnarray*}} \nc{\egae}{\end{eqnarray*}}
\nc{\ga}{\gamma} \nc{\age}{\dagger} \nc{\um}{{1\over 2}}
\nc{\al}{\alpha} \nc{\la}{\lambda} \nc{\C}{I\!\!\!C} \nc{\om}{\omega}
\nc{\R}{I\!\!R} \nc{\Abf}{\mbox{\boldmath $A$}}
\nc{\ov}{\overline} \nc{\pa}{\partial} \nc{\ug}{\; = \;}
\nc{\cent}{\centerline} \nc{\vs}{\vspace*}
\nc{\Vbf}{\mbox{\boldmath $V$}} \nc{\Ebf}{\mbox{\boldmath $E$}}
\nc{\Hbf}{\mbox{\boldmath $H$}} \nc{\imp}{\mbox{\boldmath $p$}}
\nc{\mubf}{\mbox{\boldmath $\mu$}} \nc{\Sigbf}{\mbox{\boldmath $\Sigma$}}
\nc{\sbf}{\mbox{\boldmath $s$}} \nc{\xbf}{\mbox{\boldmath $x$}}
\nc{\pibf}{\mbox{\boldmath $\pi$}} \nc{\albf}{\mbox{\boldmath $\alpha$}}
\nc{\dopsi}{\dot{\psi}} \nc{\ddopsi}{\ddot{\psi}}
\nc{\dopsib}{\dot{\overline{\psi}}} \nc{\ddopsib}{\ddot{\overline{\psi}}}
\nc{\psib}{\overline{\psi}} \nc{\dox}{\dot{x}} \nc{\ddox}{\ddot{x}}
\nc{\dddov}{{\stackrel{\ldots}{v}}} \nc{\gm}{{\ga^\mu}}
\nc{\pio}{\widehat{\pi}}  \nc{\po}{\widehat{p}}  \nc{\So}{\widehat{S}}
\nc{\dopi}{\dot{\pi}} \nc{\dop}{\dot{p}} \nc{\dov}{\dot{v}}
\nc{\qq}{\qquad \qquad} \nc{\omh}{\om_{{\rm H}}} \nc{\omz}{\om_{{\rm zbw}}}
\nc{\omp}{\om_{{\rm sp}}} \nc{\omc}{\om_{{\rm class}}}
\nc{\vecna}{\mbox{\boldmath $\nabla$}} \nc{\gabf}{\mbox{\boldmath $\gamma$}}

\begin{document}

\baselineskip 0.7cm

\cent{\bf SPIN EFFECTS ON THE CYCLOTRON FREQUENCY FOR A DIRAC ELECTRON}
\footnotetext{$^{(\dagger)}$ Work supported by INFN and by CNPq.}
\vs{0.5cm}
\begin{center}
{Giovanni SALESI\\
{\em Dipartimento di Fisica, Universit\`a  Statale di Catania,
95129--Catania, Italy; \ and\\
Istituto Nazionale di Fisica Nucleare--Sezione di Catania, Catania, Italy;\\
e-mail: {\rm Giovanni.Salesi@ct.infn.it}}\\

and\\

Erasmo RECAMI\\
{\em Facolt\`a di Ingegneria, Universit\`a statale di Bergamo, Dalmine (BG),
Italy;\\
I.N.F.N.--Sezione di Milano, Milan, Italy; \ and\\
C.C.S., State University at Campinas, Campinas, S.P., Brazil;\\
e-mail: {\rm Erasmo.Recami@mi.infn.it}}}
\end{center}

\vs{0.5 cm}

\cent{\bf Abstract}
\noindent The Barut--Zanghi (BZ) theory can be regarded as the most
satisfactory picture of a classical spinning electron and constitutes a
natural ``classical limit'' of the Dirac equation. \ The BZ model has been 
analytically studied in some previous papers of ours in the case
of free particles. \ By contrast, in this letter we consider the case
of external fields, and a previously found equation of the motion is 
generalized for a non-free spin-$\um$ particle. \  In the important case of 
a spinning charge in a uniform magnetic field, we find that its angular 
velocity 
(along its circular orbit around the magnetic field direction) is slightly
different from the classical ``cyclotron frequency'' $\omc\equiv eH/m$ 
which is expected to hold for spinless charges. \ As a matter of fact, 
the angular velocity does depend on the spin orientation.
As a consequence, the electrons with magnetic moment $\mubf$ 
parallel to the magnetic field do rotate with a frequency greater with 
respect to electrons endowed with a $\mubf$ antiparallel to $\Hbf$. \\


\newpage

\centerline{{\em PRELIMINARY VERSION}}

\section{The Barut--Zanghi model and the free-particle solutions}

In the last twenty years, renewed interest arose for classical theories of
electrons and spin$^{[1]}$; in particular, for those approaches
---as the classical Barut--Zanghi's (BZ) model of the relativistic spinning
electron$^{[2-5]}$, which we shall refer to in this paper--- that involve the
so-called {\em zitterbewegung} (zbw).$^{[6]}$ \ In the Barut--Zanghi
theory the classical electron was actually characterized, besides by the usual
pair of conjugate variables $(x^\mu, p^\mu)$, also by a second pair of conjugate
{\em classical} spinorial variables $(\psi, \ov{\psi})$, representing internal
degrees of freedom, which are functions of the (proper) time $\tau$  measured in
the center-of-mass (CM) frame; the CM frame being the one in which it is 
identically $\imp = 0$ at any instant of time.  \ Barut and Zanghi, then, 
introduced a classical Lagrangian that writes [$A^\mu$ is the electromagnetic 
potential, and the vacuum light speed $c$ is equal to 1]:
$$
\Lc=\frac{1}{2} i \la (\dot{\ov{\psi}}\psi - \ov{\psi}\dot{\psi}) +
p_\mu(\dot{x}^\mu - \ov{\psi}\ga^\mu \psi) + eA_\mu\ov{\psi}\ga^\mu \psi \; ,
\eqno{{\rm (1)}}
$$
\noindent where $\la$ has the dimension of an action, and $\psi$ and
$\ov{\psi}\equiv \psi^{\dag} \ga^0$ are ordinary
${\rm {\C}}^4$--bispinors, the dot meaning derivation with respect to $\tau$.
\ The four Euler--Lagrange equations, with $-\la=\hbar=1$, obtained by varying
$\Lc$ with respect to $\psi , \overline{\psi}, x^{\mu}, p^{\mu}$, are the 
following:
$$
\dopsi \; = \; -i\pipi\psi , \qquad \qquad \dopsib \; = \; i\psib\pipi ,
\eqno{{\rm (2a)}}
$$
$$
\dopi^{\mu} \; = \; eF^\munu\dox_{\nu}, \qquad \qquad \dox^{\mu} \; = \; \psib \ga^{\mu}\psi ,
\eqno{{\rm (2b)}}
$$
\noindent where $F^\munu$ is the electromagnetic tensor, and $\pi$ is the kinetical 
impulse
$$
\pi^\mu \equiv p^\mu - eA^\mu, \qquad \qquad  \pipi \equiv \ga_\mu \pi^\mu.
\eqno{{\rm (3)}}             
$$
Furthermore, it can be shown that quantity $\pi_\mu\psib\ga^\mu\psi \equiv \pi_\mu \dox^\mu$
results to be always a conserved quantity [assumed to be equal to $m$: see 
eq.(9a)]. \ Notice that, instead of adopting the variables $\psi$ and 
$\ov{\psi}$, we can work in terms of the following set of independent 
dynamical variables
$$
x^\mu \, , \ \pi^\mu \, ; \ v^\mu \, , S^{\mu \nu}
\eqno{{\rm (4a)}}
$$
\noindent where
$$
S^{\mu \nu} \equiv {i\over 4}\,\ov{\psi}\,[\ga^\mu, \ga^\nu]\,\psi \;
\eqno{{\rm (4b)}}
$$
\noindent is the {\em spin tensor} met in the Dirac theory; then, one gets the
following equations of the motion:

\

$\hfill{\displaystyle\left\{\begin{array}{l}
\dot{\pi}^\mu \ug eF^\munu v_\nu\\
\dot{x}^\mu \ug v^\mu\\
\dot{v}^\mu \ug 4\,S^{\mu \nu}\pi_{\nu}\\
\dot{S}^{\mu \nu} \ug v^\nu \pi^\mu - v^\mu \pi^\nu\,.
\end{array}\right.}
\hfill{\displaystyle\begin{array}{r}
(5{\rm a}) \\ (5{\rm b}) \\ (5{\rm c}) \\ (5{\rm d})\end{array}}$

\

\noindent The first equation is the well-known Lorentz equation of the motion 
for a (spinless) charged particle inside an electromagnetic (em) field 
[but for the present case of spinning electrons we have 
$\pi^\mu\neq mv^\mu$ (see below)!]; the
second one is nothing but the definition for the 4-velocity; the third equation
is formally very similar to the first one, with $v$ and $\pi$ interchanged
and the spin tensor replacing the em tensor. The last equation
expresses the conservation of the total angular momentum $J^{\mu\nu}$, sum of
the orbital angular momentum $L^{\mu\nu}$ and of $S^{\mu\nu}$:
$$
{\dot {J}}^{\mu\nu} = {\dot {L}}^{\mu\nu} + {\dot {S}}^{\mu\nu} = 0 \; ,
\eqno{{\rm (6)}}
$$
\noindent being ${\dot {L}}^{\mu\nu} = v^\mu\pi^\nu - v^\nu\pi^\mu$ from the very
definition of $L$.
\ Notice that only the first couple of equations is present in the case of
spinless particles, while the second couple of equations is directly
related to the existence of spin.
\ Starting from these Euler-Lagrange equations, in ref.[5] we worked out
the equation of the motion for the space-time coordinates $\, x^\mu$
in the particular case of {\em free} electrons ($A^\mu=0$):
$$
v \ug \frac{p}{m}-\frac{\ddot{v}}{4 m^2} \ ,
\eqno{{\rm (7)}}
$$
\noindent when assuming \ $p=$ constant \ and \ $p^2=m^2$. \         
The general solution of this equation was shown to be:$^{[2]}$
$$
v \ug \sum_{i=1}^3 v_i\,,
\eqno{{\rm (8a)}}
$$
\noindent where
$$
v_i \ug \frac{p}{m}+\left[v(0)-\frac{p}{m}\right]
\cos(\om_i\tau)+\frac{\dov(0)}{2m}\sin (\om_i\tau)
\qquad i=1,2,3\,,
\eqno{{\rm (8b)}}
$$
\noindent with$^{\#1}$
$$
\om_1=0\,; \ \qquad\qquad \om_2=2m\,; \ \qquad\qquad \om_3=-2m\,.
\eqno{{\rm (8c)}}
$$
\footnotetext{$^{\#1}$ Obviously the quantities $\om_i\equiv
({\rm d}\theta/{\rm d}\tau)_i$ can be considered as proper angular velocities
only in the CM frame or in non-relativistic frames;  otherwise one has to 
multiply the values given in eqs.(8c) by the Lorentz factor $m/\Ec$ 
(geting the so-called relativistic decreasing of the frequencies).}
A null frequency $\om_1$ is associated to a rectilinear uniform motion with 
constant velocity $v= p/m$, like in the case of a macroscopical free
body or of a non-spinning free particle. \
The particular solutions corresponding to $\om_2$ and $\om_3$, which denounce explicitly
the presence of spin, exhibit the already mentioned phenomenon of zbw.
The zbw is nothing but the spin motion, or ``internal motion" [since
it can be observed in the CM frame, where by definition $\imp=0$],  which is
expected to exist for spinning particles only. It arises because the motion 
of the electrical charge does not coincide with the motion of the CM, so that
spinning particles actually appear as extended-like objects.$^{[7]}$ \ 
In the Dirac theory, indeed, the operators velocity ($\albf$) and impulse 
(-$i\vecna$) are {\em not} parallel, in general. \ Therefore, for
motions endowed with the frequencies $\om_2$, $\om_3$), a zbw motion is to 
be added to the translational, or ``external'', motion of the CM, whose
velocity is $\, p/m$ (let us recall that the latter motion is the only one 
to be present for scalar particles). 

In other words, the 4-velocity $v$ contains in general the usual
term $p/m$ plus a term describing a motion rapidly oscillating
with the characteristic zbw angular velocities $\om=\pm 2m$. \
Let us explicitly observe that the general solution (4c)
represents a {\em helical} motion in the ordinary 3-space (a result met also
in other models and approaches$^{\#2}$ implying a zbw). \
 \footnotetext{$^{\#2}$ A physical role of the zbw has been met, and studied, 
 even in the non-relativistic framework$^{[7,8]}$ and recently extended to 
 supersymmetry and superstrings$^{[3]}$.}
Furthermore, in ref.[5] it has been found that free polarized particles
(with the spin projection $s_z$ along the $z$-axis  equal to
$\pm{1\over 2}$) are endowed with internal {\em uniform circular} motions 
around the $z$-axis. In such a way, the only {\em classical} values for 
$s_z $ corresponding to classical uniform motions in the CM frame belong to 
the discrete {\em quantum} spectrum $\pm{1\over 2}$. \
The orbit radius in the CM frame was found to be equal to $|\Vbf|/2m$
(quantity $\Vbf$ being the orbital 3-velocity), which, in the special case 
of a {\em light-like} zbw, turns out to be equal to half the Compton 
wave-length.

In the next section we want to generalize eq.(7) for the case of an electron 
in an external em field, and to write down its analytical 
solutions in the special case of a uniform magnetic field. 

\section{The motion of a classical Dirac electron in a uniform magnetic field}

Before going on, we have to assume two important constraints for physical 
consistency with the standard relativistic quantum mechanics; namely:

\

$
\hfill{\displaystyle\left\{\begin{array}{l}
\pi_\mu v^\mu \ug m\\
\pi^2 \ug m^2 + eF^\munu S_\munu\,.
\end{array}\right.}
\hfill{\displaystyle\begin{array}{r}
(9{\rm a}) \\ (9{\rm b})\end{array}}
$

\

For free electrons, the condition $p_\mu v^\mu=m$ represents the 
``classical limit" of the standard Dirac equation \ $\po_\mu\gm\,\psi \ug m\,
\psi$ \ (with $\po\equiv i\pa)$, \ as it was shown in previous 
works.$^{[2-4]}$ \ Analogously, in presence of external em fields, eq.(9a) 
may be regarded as
the ``classical limit'' of the Dirac equation in an external em field, namely
\ $\pio_\mu\gm\,\psi \ug m\,\psi$ \ (with $\pio_\mu \equiv i\pa_\mu-eA_\mu$). \ \
Notice that for spinless particles this constraint reduces to an identity;
in fact, in the absence of spin, the kinetical impulse and the velocity \ 
$v_\mu=\pi_\mu/m$ \ get parallel,  so that eq.(9a) follows by
multiplying both members by $v^\mu$.

The second condition is nothing but the ``classical limit'' of the
so-called ``second-order Dirac equation'', obtained by left-multiplying
the usual Dirac equation by $\pio_\mu\gm+m^{[9]}$. \ In fact, from
$$
(\pio_\mu\gm + m)\,(\pio_\mu\gm - m)\,\psi \ug 0\,,
$$
\noindent it follows
$$
\pio^2\,\psi \ug (m^2 + eF^\munu\So_\munu)\,\psi\,,
\eqno{{\rm (10)}}
$$
\noindent where \ $\So_\munu\equiv {i\over 4}\,[\ga_\mu, \ga_\nu]$ \ indicates the spin 
tensor operator.
\ For free ($F^\munu = 0$) spinning particles this constraint reduces
to \ $\pi^2=m^2$, \ which {\em for scalar particles} holds both in the
presence and in the absence of external fields (of course, the spin-field 
term \ $eF^\munu S_\munu$ \ is not present for spinless particles).
\ For the case of a purely magnetic field ($\Ebf=0$), that we are going to 
analyse, the constraint (9b) assumes the form:
$$
\Ec \ug \sqrt{m^2 + \pibf^2 - 2e\sbf\cdot\Hbf} \, ,
\eqno{{\rm (11)}}
$$
\noindent since for the energy $\Ec$ it holds \ 
$\Ec \equiv p^0 \equiv \pi^0+e\varphi = \pi^0$).

In the non-relativistic limit, \ $\pi^2\ll m^2$, \ from eq.(11) we easily  
get just the expected Hamiltonian for a spin-$\um$ particle in a magnetic 
field:
$$
\Ec \; \sim \; m + \frac{\pibf^2}{2m} - \frac{e
\sbf}{m}\,\cdot\Hbf\,,
\eqno{{\rm (12)}}
$$
\noindent {\em with the correct gyromagnetic ratio} $g=2$.

\

By means of a procedure analogous to the one followed for the free
case, we can now deduce the equation of the motion 
in the presence of an external em field. By deriving eq.(5c) one gets:
$$
\ddot{v}^\mu \ug 4\dot{S}^\munu\pi_\nu + 4S^\munu\dopi_\nu \ .
\eqno{(13)}
$$
\noindent Inserting eqs.(5d) and (5a) into eq.(13), and exploiting 
constraints (9a) and (9b), we finally obtain$^{\#3}$ the equation:
$$
\ddot{v}^\mu - 4m\pi^\mu + 4m^2v^\mu + 4ev^\mu F_{\la\rho}S^{\la\rho} -
4eS^\munu F_{\nu\rho}v^\rho \ug 0\,.
\eqno{(14)}
$$
\footnotetext{$^{\#3}$ In the original paper by Barut and Zanghi$^{[2]}$ a
{\em different} equation of the motion in the presence of an em field was 
deduced. Actually, those authors, besides having adopted some peculiar 
constraints, were working not in the ordinary spacetime, but in a particular
5-{\em dimensional manifold}.}
By comparison with eq.(7), holding for free particles, in the r.h.s. of the 
general equation of the motion (14) we see the appearance of two additional 
spin-field terms. The analytic solutions of eq.(14) can be easily found in 
the simple, but important, case of an external {\em uniform magnetic field} 
$\Hbf$. \ Let us take the magnetic field oriented along the $z$-axis at all
times:
$$
\Hbf \; = \; (\; 0;\; 0;\; H) \; .
\eqno{(15)}
$$
The quantum mechanical theory (the Dirac equation) entails the conservation 
of the $z$-component of the spin vector: $s_z = \pm {1 \over 2}$. \
According to the corrispondence principle between quantum mean values and
classical values, during the precession of the classical spin vector it will 
be $S^{12}=s_z$ = constant. \ Furthermore, the only nonzero components of 
the em tensor are $F^{21}=-F^{12}=H$. \
We restrict ourselves to the $xy$-plane where (by analogy with the behaviour 
of spinless charges involved by the Maxwell equations) we expect to have a
uniform circular motion of the spinning charge due to the Lorentz force.
From eq.(14) we get:

\

$
\hfill{\displaystyle\left\{\begin{array}{l}
\ddot{v}_x - 4m\pi_x + 4\left(m^2 - 3es_zH\right)\,v_x  = 0\\
\ddot{v}_y - 4m\pi_y + 4\left(m^2 - 3es_zH\right)\,v_y  = 0\,.
\end{array}\right.}
\hfill{\displaystyle\begin{array}{r}
(16{\rm a}) \\ (16{\rm b})\end{array}}
$

\

\noindent By deriving with respect to time and exploiting eq.(5a), we finally 
get:

\

$
\hfill{\displaystyle\left\{\begin{array}{l}
\dddov_x + 4\left(m^2 - 3es_zH\right)\,\dov_x - 4meHv_y = 0\\
\dddov_y + 4\left(m^2 - 3es_zH\right)\,\dov_y + 4meHv_x = 0\,.
\end{array}\right.}
\hfill{\displaystyle\begin{array}{r}
(17{\rm a}) \\ (17{\rm b})\end{array}}
$

\

\noindent This system of equations yields uniform circular motions whose
angular velocities $\om_i \ (i=1,2,3)$ are roots of the ``characteristic''
3-order algebraic equation:
$$
\om^3 - 4\left(m^2 - 3es_zH\right)\,\om + 4meH \ug 0\,.
\eqno{(18)}
$$
Even with the highest magnetic fields today experimentally achievable in
laboratory, the following condition always holds between the ``intrisic''
frequency $2m$ (that is, the zbw angular velocity for free particles)
and the ``external'' {\em cyclotron frequency} $\omh = eH/m$:
$$
\frac{eH}{m} \, \ll \, 2m\,.   
\eqno{(19)}
$$
As a consequence, we may write down the characteristic frequencies as follows:
$$
\om_1 \sim \omh\left[1 - \frac{3\omh}{m}\,s_z\right]\,; \ \qquad
\om_2 \sim 2m\left[1 - \frac{3\omh}{2m}\,s_z\right]\,; \ \qquad
\om_3 \sim -\,2m\left[1 - \frac{3\omh}{2m}\,s_z\right]\,.
\eqno{{\rm (20)}}
$$
For zero external fields, i.e., $H=0, \ \omh=0$, the above frequencies
turn out ---of course--- to be equal to the ones found for free particles, 
eq.(8c). The ``internal'' frequencies $\om_2, \ \om_3$ are related to the 
zbw motion and are substancially identical to the ones ($\pm 2m$) found for 
free particles.

The important point is that {\em the ``external'' angular velocity $\om_1$ 
results slightly different from the cyclotron frequency} $\omh$ which is 
typical of ordinary (spinless) charges in a magnetic field. The frequency 
shift is a function
of the spin vector orientation, so that {\em the spin-up and the spin-down
polarized electrons rotate with different angular velocities}:
$$
\frac{\Delta\om}{\omh} \equiv
\frac{\om_{-\um} - \om_{+\um}}{\omh} \sim 3\,\frac{\omh}{m}\,.
\eqno{(21)}
$$

Thus, we found that spinning charges with their magnetic moment
$\mubf \equiv -e\sbf/m)$
{\em parallel} to the magnetic field rotate with a frequency {\em greater} 
than the one of spinning charges having $\mubf$ antiparallel to $\Hbf$. \
The present, classical approach to the problem of a charge in a uniform
magnetic field turns out to be very suitable for describing 
{\em (non-bounded) electrons
performing large orbits in vacuum}, which behave as classical bodies.

In fact, in cyclotrons, magnetic rings or bottles, and non-linear 
accelerators, the angular velocity is usually assumed to be equal to the 
classical value $\omh$ for both the possible polarizations $s_z=\pm\um$.
The small frequency shift predicted, on the contrary, by eq.(21) could be 
observed by means of an {\em ad hoc} experiment, in which a device 
measuring $s_z$ (or, anyway, interacting in a {\em different} way with the 
different polarizations) is placed at a point along the orbit.
For eq.(21), a particle beam, initially containing both the spin
components, will progressively spread along the orbital motion: so that the
spin-down particles (which rotate faster) will slowly separate from the
spin-up particles. 

For example, for a cyclotron orbit with a diameter $d=1\,$m 
only and a magnetic field $H=3.4\cdot10^{-4}\,$T, we obtain, for electrons, 
the orbital speed $v={1\over 10}c$ (which may be still considered as 
non-relativistic, so that the relativistic frequency decrease is negiligible), 
and the angular velocity $\om\sim6\cdot10^7\,$Hz. \ As a consequence, the 
frequency shift $\Delta\om$ will be about $1.34\cdot10^{-6}\,$Hz, which 
implies a phase difference of $2\pi$ (corresponding to one full orbit) 
in about $7.5\cdot10^6\,$s. \ After a lapse of time $\Delta t = 40$ minutes, 
the spin-down electrons are expected to precede the spin-up ones by a distance
$\Delta l =1\,$mm. \ As a consequence, the spin-down particles will interact 
with a suitable detector at a time $\Delta\tau = 4.27\cdot10^{-10}\,$s 
earlier than the spin-up particles.

\

\

{\bf Acknowledgements}

\noindent The scientific collaboration of G.Andronico, G.G.N.Angilella, 
A.Bonasera, M.Borrometi, L.Bosi, G.Cavalleri, L.D'Amico, S.Esposito,
A.Gigli Berzolari, G.Di Lorenzo, S.Dimartino, C.Dipietro, P.Falsaperla, 
G.Fonte, H.Hern\'andez F., L.C.Kretly, L.Lo Monaco, S.Lo Nigro, G.Lucifora, 
G.Marchesini, E.C.Oliveira, M.Pignanelli, R.Petronzio, S.Sambataro, R.Turrisi, 
J.Vaz, D.Wisnivesky and M.Zamboni-Rached is also acknowledged. \ For the kind, 
active cooperation special thanks are due to Rosellina Salesi, G.Giuffrida 
and M.T.Vasconselos.\\

\

\end{document}